\documentclass[12pt]{article}
\usepackage[dvips]{graphicx}
\usepackage[tight]{subfigure}
\def\be{\begin{equation}}
\def\ee{\end{equation}}
\def\bea{\begin{eqnarray}}
\def\eea{\end{eqnarray}}
\def\beaN{\begin{eqnarray*}}
\def\eeaN{\end{eqnarray*}}
\def\ed{\end{document}}
\def\bit{\begin{itemize}}
\def\eit{\end{itemize}}

\def\sig{\sigma}

\def\alf{\alpha}

\def\di{\partial}

\def\~{\tilde}
\def\lag{{{\cal L}}}
\def\m{\label}
\def\l{\left}
\def\r{\right}

\def\goto{\rightarrow}
\def\Bar{\overline}
\def\const{\rm const}
\def\diag{\rm diag}

\def\sA{\stackrel{\bullet}{A}}

\def\sS{\stackrel{\bullet}{S}}

\def\sJ{\stackrel{\bullet}{J}}

\def\sK{\stackrel{\bullet}{K}}

\def\sL{\stackrel{\bullet}{\lag}}
\def\cN{\stackrel{\circ}{\nabla}}
\def\cA{\stackrel{\circ}{A}}

\def\scJ{\stackrel{\bullet}{\cal J}}
\def\sK{\stackrel{\bullet}{K}}

\def\srho{{\stackrel{\bullet}{\rho}}{}}

\usepackage{authblk}

\begin{document}

\title{ \bf  A moving black hole in TEGR as a moving matter ball}
\author[1,2]{E. D. Emtsova\thanks{Electronic address: \texttt{ed.emcova@physics.msu.ru}}}

\author[1]{A. N. Petrov\thanks{Electronic address: \texttt{alex.petrov55@gmail.com}}}

\affil[1]{Sternberg Astronomical institute, MV Lomonosov State university  \protect\\ Universitetskii pr., 13, Moscow, 119992,
Russia}
\affil[2]{Physical Faculty, M.V.Lomonosov Moscow State University, Moscow 119991, Russia}

\date{\small \today}
\maketitle

\begin{abstract}
    Possibilities of the covariant with respect to both coordinate and local Lorentz
transformations formalism developed earlier in the framework of Teleparallel Equivalent of General Relativity (TEGR) are studied. The formalism is applied to a solution for a moving with constant velocity (with respect to distant static observers)
Schwarzschild black hole. Coordinate and Lorentz invariant global conserved mass
and momentum are constructed. The acceptable results are obtained in spite of the
solution under consideration has no, at least, Killing vectors of space displacements.
Calculations are quite analogous to calculating the mass and momentum of a moving
matter ball in Minkowski space, and this analogy is used essentially.
\end{abstract}

\section*{Introduction}
\m{Introduction}

Teleparallel gravity  including TEGR (Teleparallel Equivalent of General Relativity) has been developing very actively over the past few decades \cite{Aldrovandi_Pereira_2013,REV_2018,REV_2021}. In many researches in teleparallel gravity, black hole  solutions are the most widely used as models to test various developing formalisms. Among them, Schwarzschild black hole solution is considered in detail  \cite{M_2M,Maluf_2005,Maluf,Maluf0704,Obukhov+} and is being  used  to calculate the mass of a black hole as a conserved global energy or derive energy density of the gravitational field measured by an observer. Many of approaches, see for example \cite{Maluf0704, Obukhov+}, lead either to non-covariant with respect to coordinate transformations, or non-invariant with respect to local Lorentz rotations quantities.

 Nevertheless, a fully covariant formalism can be developed with the use of the Noether theorem. Thus, in \cite{Obukhov_2006,Obukhov:2006ge,Obukhov_Rubilar_Pereira_2006} fully covariant conserved quantities are constructed in the differential form presentation, whereas in \cite{EPT_19, EPT_2020} they are constructed in the more popular tensorial presentation. Namely, the formalism of \cite{EPT_19,EPT_2020} is applied in the present paper. To achieve the covariance of both types, at least, one needs to introduce a non-dynamical quantity representing the inertial effects - the inertial spin connection (ISC) that is not determined by the theory itself. In \cite{EPT_19, EPT_2020}, the principle of ``turning off'' gravity is used just to determine ISC. It is based on the fact that in the absence of gravity, only inertial effects can remain. In this case, the curvature tensor vanishes and then Levi-Civita spin connection (L-CSC) expresses only inertial effects and should be equal to the ISC. Applying the Noether approach and considering the invariance under diffeomorphisms induced by an arbitrary smooth vector field $\xi$, one needs to choose $\xi$ in a physically meaningful way. They can be Killing vector fields of the reference geometry, proper vectors of observers, etc.


For the best of our knowledge, the Schwarzschild solution was not considered in TEGR
as a moving black hole in any applications. Thus, the first goal here is to close this gap
and calculate the global conserved energy and momentum for the moving black hole \cite{HTP}
with making the use of the method \cite{EPT_19,EPT_2020}. However, calculating the total mass and total
momentum is not an end in itself. Indeed, the solution \cite{HTP} is asymptotically flat and these
quantities are known and can be obtained by other related methods in general relativity \cite{Petrov_book}.
Only we would want to demonstrate possibilities of the covariant formalism \cite{EPT_19,EPT_2020} and its
advantages. It is the second goal. It turns out that these calculations are quite analogous to
calculating the mass and momentum of a moving matter ball in Minkowski space. We use
this analogy greatly.

\section{Covariant conserved quantities in TEGR}
\m{Second}

Let us outline the covariant formalism \cite{EPT_19,EPT_2020}. The gravitational Lagrangian of TEGR:
\be
\sL = \frac{h}{2\kappa} \l({\sK}{}^\rho{}_{\mu\nu} {\sK}_\rho{}^{\nu\mu} - {\sK}{}^\nu{}_{\rho\nu} {\sK}{}^{\mu\rho}{}_\mu \r) \,
\m{lag}
\ee
is equivalent to the Hilbert Lagrangian up to a divergence. Dynamical variables in TEGR are components of the tetrad field $h^a{}_\rho$, which are connected with metric by $g_{\mu\nu}=\eta_{ab} h^{a}{}_{\mu} h^{b}{}_{\nu}$ and $h \equiv \det h^a{}_\rho$, where $\eta_{ab}$ is the Minkowski metric; Greek indexes are spacetime components, Latin indexes  $a,b,c,\ldots$ are tetrad components,  Latin indexes $i,j,k,\ldots$ are space components.  The contortion tensor is defined as $\sK{}^a{}_{b\nu}= \sA{}^a{}_{b\nu}-\cA{}^a{}_{b\nu}$, where ${\cA}{}^a{}_{b \nu} = - h_b {}^\rho \cN_\nu h^a {}_\rho$ is the L-CSC with $\cN_\nu$ Levi-Civita covariant derivative, ${\sA}{}^a{}_{b \nu}$ is the teleparallel ISC. Tetrad indexes are replaced by spacetime indexes and inversely by contracting with $h^{a}{}_{\mu}$ or $h_{a}{}^{\mu}$, for example, $\sK{}^\rho{}_{\mu \nu}=\sK{}^a{}_{b\nu} h_a {}^\rho h^b {}_\mu$.

 Simultaneous transformations of tetrad and ISC under local Lorentz rotations are:
\begin{equation}\label{transformation_tet}
   h'{}^a {}_{\mu} = \Lambda {}^a {}_b  h^b{}_\mu,
\end{equation}
\begin{equation}\label{transformation_ISC}
\sA{}'{}^a {}_{b \mu}=\Lambda {}^a {}_c \sA{} {}^c {}_{d \mu} \Lambda {}_b {}^d   + \Lambda {}^a {}_c \partial_\mu \Lambda {}_b {}^c,
\end{equation}
where $\Lambda^b{}_d(x)$ is a matrix of a local Lorentz transformation. The L-CSC  ${\cA}{}^a{}_{b \nu}$ is transformed analogously to (\ref{transformation_ISC}). Then it is evidently that $\sK{}^\rho{}_{\mu \nu}$ is invariant under local Lorentz transformations.
Because $\sA{} {}^c {}_{d \mu}$ represents the inertial effects of the tetrad it can be suppressed by  (\ref{transformation_ISC}) with appropriate $\Lambda {}^a {}_c$ \cite{EPT_19,EPT_2020}.

Considering the invariance under diffeomorphisms induced by an arbitrary vector field $\xi$, one derives for the action (\ref{lag}) conservation law for the current  ${\scJ}{}^\alf(\xi)$:
  \be
\di_\alf {\scJ}{}^\alf(\xi) = \cN_\alf {\scJ}{}^\alf(\xi) =0.
\m{DiffCL_2}
\ee
The current itself is expressed through the superpotential:
\begin{equation}\label{DiffCL_1}
{\scJ}{}^\alf(\xi)=\partial_\beta {\scJ}{}^{\alf\beta}(\xi)=\cN_\beta {\scJ}{}^{\alf\beta}(\xi).
\end{equation}
Noether's current $ {\scJ}{}^\alf(\xi)$ is the vector density of the weight +1, Noether's superpotential $ {\scJ}{}^{\alf \beta}(\xi)$ is the antisymmetric tensor density of the weight +1, in TEGR it is
\begin{equation}\label{super}
 {\scJ}{}^{\alf\beta}(\xi)=
  \frac{h}{\kappa}\sS_\sigma{}^{\alpha\beta}\xi^\sigma ; \qquad{\sS}_\sig{}^{\alpha\beta}=
     \sK{}^{\alpha\beta} {}_\sig + \delta^\beta_{\sigma} \sK{}^{\theta \alf} {}_{\theta} - \delta^\alf_{\sigma}  \sK{}^{\theta \beta} {}_{\theta}.
\end{equation}
Both ${\scJ}{}^{\alf}(\xi)$ and ${\scJ}{}^{\alf\beta}(\xi)$ are locally Lorentz invariant.
In the case of spherical symmetry, when $r=x^1$, conservation laws (\ref{DiffCL_2}) and (\ref{DiffCL_1}) allow us to construct a conserved charge:
\begin{equation}\label{noetcharge}
    {\cal P}(\xi) = \int_\Sigma d^3x  {\scJ}{}^0(\xi) = \oint_{\di\Sigma} d^2 x{\scJ}{}^{01}(\xi) =    \frac{1}{\kappa}
\oint_{\partial \Sigma}d^2x \, h\, \sS_\sigma {}^{01}\xi^\sigma  ,
\end{equation}
where $\Sigma$ is a hypersurface of constant time $t=x^0=\const$, $\di\Sigma$ is a boundary of $\Sigma$, and can be considered both at finite $r=r_0$ and at $r\goto \infty$.

\section{A moving matter ball in Minkowski space}
\m{Third}

From the start, because the model of moving black hole is similar to moving spherically distributed matter in Minkowski space, it is instructive to consider such a matter in a flat spacetime with metric:
\begin{equation}\label{Minkmet}
    ds^2=-dt^2+dx^2+dy^2+dz^2.
\end{equation}
 Here, we denote these coordinates as $(t,x,y,z) = (x^0, x^i) = (x^\alpha)$, where $i=1,2,3$. To turn this spacetime into a reference frame, we add it by static observers which  have  proper vectors $\xi^\alf=(-1, 0, 0, 0)$.
 Let the matter in the Minkowski space have energy-momentum tensor $\Theta^\alpha {}_\beta$, which is differentially conserved, $\di_\alf\Theta^\alpha {}_\beta =0$. Then the current ${{\cal J}}{}^{\alpha} = \Theta^\alpha {}_\beta \xi^\beta$ is conserved, $\di_\alf{{\cal J}}{}^{\alpha} = 0$, as well, and its components present the  energy density  ${{\cal J}}{}^{0} = \Theta^0 {}_0\xi^0$ and the momentum density ${{\cal J}}{}^{i} = \Theta^i {}_0 \xi^0$ measured by the introduced above observers. One recognizes that the definition of the current ${\scJ}{}^{\alf}(\xi)$ in (\ref{DiffCL_2}) and (\ref{DiffCL_1}) is a generalization to TEGR of the simplest definition in Minkowski space, and the components of ${\scJ}{}^{\alf}(\xi)$ have the analogous interpretation for observers with proper vectors $\xi$.

Assuming the static spherically symmetric distribution of matter, one has for the current
\begin{equation}\label{curr_stat_matter}
    {{\cal J}}{}^{\alpha}_s = \l[\rho  (r), ~0, ~0, ~0\r],
\end{equation}
where $\rho(r) =  {{\cal J}}{}^{0}_s (r) =\Theta^0 {}_0 (r) \xi^0$ is just the energy density, $r^2 \equiv x^2+y^2+z^2$ and
\begin{equation}\label{coordinates}
        x=r \sin \theta  \cos \phi ;~~
        y=r \sin \theta \sin \phi;~~
        z=r \cos \phi .
\end{equation}
 Suppose that matter distributed on the hypersurface $\Sigma$ of a constant time is bounded by $\di\Sigma$ that is a sphere $r=r_0$. Then the mass/energy of such a matter ball can be calculated  as
\begin{equation}\label{mattermassenergy}
  E_s = {\int\!\!\int\!\!\int_\Sigma} dx dy dz {{\cal J}}{}^{0}_s  (r) ={\int\!\!\int\!\!\int_\Sigma} dx dy dz \rho(r) = \int^{r_0}_0\!\!\int^\pi_0\!\!\int^{2\pi}_0 drd\theta d\phi\, \rho(r)r^2\sin\theta  = {\cal M}.
\end{equation}

Let us assume that an absolutely identical matter ball is moving relatively to the frame $\{{x}^\alf \}$ connected with (\ref{Minkmet}) with the constant velocity $v$ along the axis $0\div x$ . The proper coordinates of the moving object are connected with those in (\ref{Minkmet}) by the Lorentz transformation:
\begin{equation}\label{movingcoord}
    \Bar{t}=\gamma (t-v x);~~  \Bar{x}=\gamma (x-v t);~~ \Bar{y}=y; ~~\Bar{z}=z,
\end{equation}
where $\gamma\equiv(1-v^2)^{-\frac{1}{2}}$. In analogy with the reference frame $\{ x^\alf \}$ determined by (\ref{Minkmet}) the moving ball has a {\em proper} (its own) reference frame $\{ \Bar{x}^\alf \}$.

To illustrate the below calculations related to a moving matter ball we consider a simpler case. Let the moving sphere be filled by $N$ point particles with masses $m$ and static in the proper frame $\{ \Bar{x}^\alf \}$, thus the total mass in $\{ \Bar{x}^\alf \}$ is ${\cal M}_s=Nm$. After that, let us find the mass of such a moving object in the frame  $\{ x^\alf \}$.   First, the moving sphere undergoes relativistic compression and its volume decreases $\gamma$ times. Second, by effects of special relativity, energy and momentum of each particle with mass $m$ becomes $\gamma m$, and $\gamma v m$. At last, third, because the number of particles $N$ is conserved the concentration of particles increases in $\gamma$ times. For this simplest model it is clear that the first and third factors are compensated. Then, the evident result for the total mass and momentum of the moving object becomes ${\cal M}_m = N(\gamma m) = \gamma {\cal M}_s$ and ${\cal P}^1_m = N(\gamma v m) = \gamma v{\cal M}_s$.

The above logic is working in the case of continuous matter distribution as well. Thus in the proper frame $\{ \Bar{x}^\alf \}$ of the moving ball the current has the form:
\begin{equation}\label{curr_mov_matter_proper}
        {{\cal J}}{}^{\Bar{\alpha}}_s = \l[\rho(\Bar{r}), ~0, ~0, ~0\r],
\end{equation}
where $\Bar{r}^2 = \Bar{x}^2+\Bar{y}^2+\Bar{z}^2$ and $\rho  (\Bar{r})$ is the same function like in (\ref{curr_stat_matter}).
Now, let us transform from the frame $\{ \Bar{x}^\alf \}$ to the frame $\{ {x}^\alf \}$. The first factor of the relativistic compression of the sphere is to be taken into account in boundaries of integration. By the second factor, the components of the vector (\ref{curr_mov_matter_proper}) after Lorentz transformations (\ref{movingcoord}) become
\begin{equation}\label{curr_mov_matter_transformed}
     {{\cal J}}{}^{{\alpha}}_m = \l[\gamma \rho(\Bar{r}),\gamma v \rho(\Bar{r}), ~0, ~0\r]
\end{equation}
in the frame $\{ x^\alf \}$ in coordinates $(t,x,y,z)$, where $\Bar{r}^2 = \gamma^2 ({x} - vt)^2+{y}^2+{z}^2$. By the third factor, due to the relativistic compression the components of (\ref{curr_mov_matter_transformed}) have to be multiplied by $\gamma$ under the integration in the compressed boundaries.

Thus, to obtain the total mass of the moving matter ball one integrates
\begin{equation}\label{energy_matter}
  {E}_m= {\int\!\!\int\!\!\int_\Sigma} d {x} d {y} d {z}  \l(\gamma {{\cal J}}{}^{0}_m (\Bar{r}) \r)= \gamma {\int\!\!\int\!\!\int_\Sigma} dx' dy dz \rho (r')= 4\pi\gamma \int^{r_0}_0 dr'\,r'^2 \rho(r')  =   \gamma {\cal M},
\end{equation}
where the boundary ${\di\Sigma}$ of $\Sigma$ is defined as $\gamma^2 x^2+y^2+z^2=r_0^2$  calculated at the moment $t=0$ without the loss of generality. After the simple redefinition $x'=\gamma x$ one has $x'^2+y^2+z^2=r'^2$ and the boundary ${\di\Sigma}$ is defined as usual $r'=r_0$, thus the last integration in (\ref{energy_matter}) repeats exactly (\ref{mattermassenergy}). The same, following the analogy with the example of point particles, where the total momentum is obtained by a simple summation, the total momentum of the moving matter ball is obtained by integration of a momentum density ${{\cal J}}{}^{1}_m (\Bar{r})$ multiplied by $\gamma$
\begin{equation}\label{momentum_matter}
  {P}^1_m= {\int\!\!\int\!\!\int_\Sigma} d {x} d {y} d {z}  \l(\gamma {{\cal J}}{}^{1}_m (\Bar{r}) \r)= \gamma v{\int\!\!\int\!\!\int_\Sigma} dx' dy dz \rho (r')= 4\pi\gamma v \int^{r_0}_0 dr'\,r'^2 \rho(r')  =   \gamma v{\cal M}.
\end{equation}
First, the results (\ref{energy_matter}) and (\ref{momentum_matter}) are in the full correspondence with the conclusions of special relativity. Second, namely the above prescription will be used to calculate the global mass and momentum of the moving black hole because of the covariant formalism of \cite{EPT_19, EPT_2020} allows us to realize this program.

\section{A moving black hole in TEGR}
\m{Forth}

Analogously to the matter ball in Minkowski space we begin to study the Schwarzschild solution in a static presentation. Following  \cite{HTP}, we take the Schwarzschild metric in isotropic coordinates:
\begin{equation}\label{isomet}
ds^2=- \alpha^2 (r) dt^2+\psi^4 (r) (dx^2+dy^2+dz^2),
\end{equation}
where $\alpha (r) \equiv (1-\frac{M}{2r})/(1+\frac{M}{2r})$,  $\psi (r) \equiv 1+\frac{M}{2r}$ and again $x^2+y^2+z^2=r^2$.
The most convenient is to choose the tetrad in diagonal form:
\begin{equation}\label{isotet}
h^a {}_\mu = \diag \l[\alpha (r), \psi^2 (r), \psi^2 (r), \psi^2 (r)  \r].
\end{equation}

Non-zero components of L-CSC ${\cA}{}^a{}_{b \nu} = - h_b {}^\rho \cN_\nu h^a {}_\rho$ calculated for (\ref{isomet}) and (\ref{isotet}) are:
\begin{equation}\label{isoLCSC}
        \cA{}^{\hat{0}} {}_{\hat{i} 0}=- \cA{}^{\hat{i}} {}_{\hat{0} 0} = \frac{M x^i}{r^3} \frac{1}{\psi^4 (r)};~~
              \cA{}^{\hat{i}} {}_{\hat{k} i}=- \cA{}^{\hat{k}} {}_{\hat{i} i} = \frac{M x^k}{r^3} \frac{1}{\psi (r)},
\end{equation}
where the indexes with ``hat'' are tetrad components and indexes without ``hat'' are spacetime components; here, $x^i \equiv (x^1, x^2, x^3) \equiv (x,y,z)$. ``Turning-off gravity'' in L-CSC (\ref{isoLCSC}) is provided by the condition $M \goto 0$ what leads to vanishing L-CSC giving for all the components of ISC, $\sA{}^{a} {}_{b \mu}=0$.
 Within the definitions in \cite{Aldrovandi_Pereira_2013,Obukhov+,EPT_19}, a tetrad with a related zero ISC is called a ``proper tetrad''. For the L-CSC (\ref{isoLCSC}) and zero ISC the formulae (\ref{super}) give the teleparallel superpotential $\sS{}_{\sig} {}^{\alf\beta}$, non-zero components of which are:
 \begin{equation}\label{isosupcoord}
        \sS{}_{{0}} {}^{{0} i}=- \sS{}_{{0}} {}^{{i} 0} =
        \frac{2 M x^i}{r^3} \frac{1}{\psi^5 (r)};~~
  \sS{}_{{i}} {}^{{i} k}=- \sS{}_{{i}} {}^{{k} i} =
        \frac{M^2 x^k}{2 r^4} \frac{1}{\alpha (r) \psi^6 (r)}.
\end{equation}

To calculate the total mass of the Schwarzschild black hole, it is more convenient to take the spherical coordinates. Therefore, let us provide the standard coordinate transformation (\ref{coordinates}) after that the metric (\ref{isomet}) acquires the form:
\begin{equation}\label{isometR}
ds^2=- \alpha^2 (r) dt^2+\psi^4 (r) \l[dr^2+ r^2 (d\theta^2 + \sin^2\theta d\phi^2) \r].
\end{equation}
We take again the convenient diagonal tetrad for (\ref{isometR}):
\begin{equation}\label{isotetR}
h^a {}_\mu = \diag \l[\alpha (r), \psi^2 (r),  r \psi^2 (r), r \psi^2 (r)\sin \theta \r].
\end{equation}
For the metric (\ref{isometR}) and tetrad (\ref{isotetR}), the non-zero components of L-CSC are
\begin{equation}\label{isoLCSCR}
    \begin{array}{cccc}
        \cA{}^{\hat{0}} {}_{\hat{1} 0} = \cA{}^{\hat{1}} {}_{\hat{0} 0} = \frac{M}{r^2} \frac{1}{\psi^4 (r)};~
\cA{}^{\hat{1}} {}_{\hat{2} 2}=-\cA{}^{\hat{2}} {}_{\hat{1} 2} = \frac{M}{r} \frac{1}{\psi (r)}-1;\\
\cA{}^{\hat{1}} {}_{\hat{3} 3}=-\cA{}^{\hat{3}} {}_{\hat{1} 3} = -  \alpha (r)\sin \theta;~
\cA{}^{\hat{2}} {}_{\hat{3} 3} =-\cA{}^{\hat{3}} {}_{\hat{2} 3} = -\cos \theta ,
    \end{array}
\end{equation}
where now $x^i \equiv (x^1, x^2, x^3) \equiv (r, \theta, \phi)$.

``Turning off'' gravity by $M \goto 0$ in (\ref{isoLCSCR}) gives ISC, non-zero components of which are:
\begin{equation}\label{isoISCR}
        \sA{}^{\hat{1}} {}_{\hat{2} 2}=-\sA{}^{\hat{2}} {}_{\hat{1} 2} = -1;~~
\sA{}^{\hat{1}} {}_{\hat{3} 3}=-\sA{}^{\hat{3}} {}_{\hat{1} 3} = -\sin \theta;~~
\sA{}^{\hat{2}} {}_{\hat{3} 3}=-\sA{}^{\hat{3}} {}_{\hat{2} 3} = -\cos \theta.
\end{equation}
Then, formulae (\ref{super}) for the L-CSC (\ref{isoLCSCR}) and ISC (\ref{isoISCR}) give non-zero components of $\sS{}_{\sig} {}^{\alf\beta}$:
\begin{equation}\label{isosupcoordR}
     \sS{}_{{0}} {}^{{0} 1}=-\sS{}_{{0}} {}^{{1} 0} = -\frac{2M}{r^2} \frac{1}{\psi^5 (r)};~
\sS{}_{{2}} {}^{{1} 2}=-\sS{}_{{2}} {}^{{2} 1}=\sS{}_{{3}} {}^{{1} 3}=-\sS{}_{{3}} {}^{{3} 1}  = -\frac{M^2}{2r^3} \frac{1}{\alpha (r) \psi^6 (r)}.
\end{equation}
After applying the transformations (\ref{transformation_tet}) and (\ref{transformation_ISC}) to the tetrad (\ref{isotetR}) and ISC (\ref{isoISCR}),  where
\begin{equation}\label{lambdabhxyz}
 \Lambda^a {}_b =  \left(
\begin{array}{cccc}
 1 & 0 & 0 & 0 \\
 0 & \sin \theta  \cos \phi  & \cos \theta  \cos \phi  & -\sin \phi  \\
 0 & \sin \theta  \sin \phi  & \cos \theta  \sin \phi  & \cos \phi  \\
 0 & \cos \theta  & -\sin \theta  & 0 \\
\end{array}
\right),
\end{equation}
 the obtained ISC vanishes and the obtained tetrad becomes a proper tetrad. This proper tetrad after the coordinate transformations (\ref{coordinates}) becomes equal to the proper tetrad (\ref{isotet}) and the superpotential (\ref{isosupcoordR}) goes to (\ref{isosupcoord}). In the framework of the covariant formalism \cite{EPT_19,EPT_2020}, arbitrary pairs of tetrads and ISCs  connected by smooth coordinate and/or local Lorentz transformations present the same {\em gauge} in terminology of \cite{EKPT_2021}, and can be used in calculations to obtain the same result. Here, in calculations, we use the gauge presented by the
 pair of tetrad (\ref{isotet}) and zero ISC (or by the pair of tetrad (\ref{isotetR}) and ISC (\ref{isoISCR})) only.\footnote{In spite of evident advantages of covariant formalism the ambiguity related to constructing conserved quantities remain. The reason is that ISC cannot be chosen in an unique way even by the ``turning off'' gravity principle. Recently, this problem has been studied in detail in \cite{EKPT_2021,EKPT_2021a}.}

To calculate the global mass/energy of a black hole, it is necessary to determine the observers in the same way as it was done in the Minkowski space. A spacetime with metric (\ref{isomet}), or (\ref{isometR}), and with the 4-vectors of static observers
\begin{equation}\label{staticobserver}
    \xi^\sigma = \l[-\alpha^{-1} (r), ~0, ~0, ~0\r]
\end{equation}
presents a static reference frame $\{x^\alf\}$.
Then, (\ref{super}) with (\ref{isosupcoordR}) gives the unique non-zero component of the Noether superpotential
\begin{equation}\label{isonetsup}
    \scJ{}^{{0} 1}_s=-\scJ{}^{{1} 0}_s =  {2{\kappa}^{-1}M}  \psi (r)  \sin \theta  ;
\end{equation}
and (\ref{DiffCL_1}) gives the Noether current in the form:
\begin{equation}\label{isonetcur}
  \scJ{}_s^\alpha =    \l[{2{\kappa}^{-1}M}  \psi' (r)  \sin \theta  ,~ 0, ~0, ~0 \r].
\end{equation}
 Because $\scJ{}_s^\alpha$ is a vector density, see (\ref{DiffCL_2}) and (\ref{DiffCL_1}), the energy density in (\ref{isonetcur}) presented in spherical coordinates can be rewritten as $\scJ{}_s^0 = {{\srho}}(r)r^2\sin\theta$, where  $ {{\srho}}(r)$ is the energy density in the Cartesian coordinates of (\ref{isomet}). Thus, substituting  (\ref{isonetcur}) and (\ref{isonetsup}) into (\ref{noetcharge}) we get
\begin{equation}\label{E=M}
E_s   = \lim_{r_0\goto\infty}{\int\!\!\int\!\!\int_\Sigma} dx dy dz {{\srho}}(r) =\lim_{r_0\goto\infty}{\int\!\!\int\!\!\int_\Sigma} dr d\theta d\phi {{\sJ}}{}^{0}_s = \lim_{r_0\goto\infty} \oint_{\di\Sigma} d\theta d\phi {{\sJ}}{}^{01}_s  =  M,
\end{equation}
where again the boundary $\di\Sigma$ of $\Sigma$ presents a sphere $r=r_0$, and then one takes the limit ${r_0\goto\infty}$.
The result (\ref{E=M}) can be interpreted as the global mass of the black hole, since at $r_0 \goto \infty$ the 4-vector (\ref{staticobserver}) asymptotically tends to the timelike Killing vector $\xi^\alf=(-1,~0,~0,~0)$. If the charge (\ref{E=M}) is calculated at finite $r=r_0$, it can be interpreted as the energy measured by observers resting at $r=r_0$. The acceptable result (\ref{E=M}) shows us that our choice of the gauge in the terminology \cite{EKPT_2021,EKPT_2021a} just corresponds to the problem of calculating the global mass.

To describe the moving black hole the authors \cite{HTP} apply the transformation (\ref{movingcoord}) to the {\em barred} metric (\ref{isomet}) and obtain the metric of the moving black hole in the frame $\{ {x}^\alf \}$:
\begin{equation}\label{movemet}
    d {{s}}^2 = - {\Bar{\alf}}^2 d {{t}}^2 +\gamma^2 (\psi^4-\alf^2 v^2)(d {{x}} + {\Bar{\beta}} d {{t}}^2)^2 + \psi^4 (d {{y}}^2+ d {{z}}^2).
\end{equation}
Here ${\Bar{\alf}}=\alpha \gamma^{-1} (1-\alpha^2 v^2/\psi^4)^{-1/2}$; ${\Bar{\beta}}= -v (1-\alpha^2/\psi^4) (1-\alpha^2 v^2/\psi^4)^{-1}$,  $\alpha = \alpha (\Bar{r})$ and   $\psi = \psi (\Bar{r})$, where ${\Bar{r}}^2 =\gamma^2 ({{x}} - vt)^2+{{y}}^2+{{z}}^2$. Let us turn to the {\em proper} reference frame of the moving black hole $\{\Bar{x}^\alf \}$ defined by the {\em barred} metric (\ref{isomet}) and related observers analogous to (\ref{staticobserver}). Repeating all the steps done for the static black hole and preserving the static gauge which is obtained relatively to the black hole, we get in the coordinates $(\Bar t, \Bar x, \Bar y, \Bar z)$:
\begin{equation}\label{isonetcurbh}
      \scJ{}_s^{\Bar{\alpha}} (\Bar{r})=    \l[\srho{} (\Bar{r}) ,~ 0, ~0, ~0 \r],
\end{equation}
where the dependence $\srho{} (\Bar{r})$ is exactly the same as defined for (\ref{isonetcur}) and ${\Bar{r}}^2 ={{\Bar x}}^2+{{\Bar y}}^2+{{\Bar z}}^2$. We emphasize that in the frame $\{\Bar{x}^\alf \}$ we, of course, repeat the result (\ref{E=M}): $\Bar E_s=M$ for the global mass.

Because the gauge is already chosen, and the solution (\ref{movemet}) is obtained from the barred metric (\ref{isomet}) with the use of (\ref{movingcoord}), the covariant formalism \cite{EPT_19,EPT_2020} allows us to transform the components of the current (\ref{isonetcurbh}) with the use of (\ref{movingcoord}) to the frame $\{{x}^\alf \}$:
\begin{equation}\label{movenetcurbh}
          \scJ{}_m^{\alpha } (\Bar{r})=    \l[\gamma \srho{} (\Bar{r}) ,~\gamma v  \srho{} (\Bar{r}) ,~ 0, ~0 \r].
\end{equation}
 Formally (\ref{movenetcurbh}) coincides with the current (\ref{curr_mov_matter_transformed}) for a matter ball in Minkowski space. Likewise, the integration for the components of (\ref{movenetcurbh}) actually repeats the integration in (\ref{energy_matter}) and (\ref{momentum_matter}). The only difference is that according to (\ref{noetcharge}) one can go to the surface integration like in (\ref{E=M}), and then take the limit $r'={r}_0 \goto \infty$. Finally, we get the global mass and momentum for the moving black hole:
\begin{equation}\label{total_bh_mass_momentum}
    E_m = \gamma M; \qquad P^1_m = \gamma v M.
\end{equation}

\section{Conclusion}
\m{Fifth}

Let us discuss the results. First, the transformation (\ref{movingcoord}) applied to (\ref{isomet}) presents the global Lorentz boost for distant observers. Thus, the results (\ref{total_bh_mass_momentum}), presenting the global (at infinity) conserved quantities, can be interpreted as quite acceptable ones.

Second, the result (\ref{E=M}) was obtained by choosing (\ref{isomet}), or (\ref{isometR}), with observers (\ref{staticobserver}). The
same result could be obtained with the use of the Killing vector  $\xi^\alf = (-1,~0,~0,~0)$ instead
of (\ref{staticobserver}). However, following this logic, to calculate the momentum density, a space-like
displacement Killing vector would be needed, but it does not exist for the solution under
consideration. Here, following the analogies in Minkowski space we interpret components of
the conserved current for a static black hole as components of the energy-momentum vector
when energy density and momentum density are measured by local observers. Components
of the current for a moving black hole are obtained by related transformation (\ref{movingcoord}) with the
following integration and final results (\ref{total_bh_mass_momentum}).

Third, in  \cite{EKPT_2021a} it was found that for the same solution, the ``turning-off gravity'' in different frames can lead to different ISCs, including those which give unacceptable results. Here
we have used the only gauge in which tetrad (\ref{isotet}) is a proper tetrad, i.e. associated with
vanishing ISC, and this gauge leads to physically meaningful results (\ref{E=M}) and (\ref{total_bh_mass_momentum}). Nevertheless, the problem of determining different gauges for the solution (\ref{movemet}) is interesting, and
its study is carried over to the future.

\bigskip

{\bf Acknowledgments.} The authors are grateful to Alexey Toporensky for discussions and comments. AP thanks Stephen Lau and Brian Pitts for the help with literature. This research has been supported by the Interdisciplinary Scientific and Educational School of Moscow State University ``Fundamental and Applied Space recearch''.

\end{document}